\documentclass[useAMS,usenatbib]{mn2e}

\title[Complex emission patterns]{Complex emission patterns: fluctuations and bistability
of polar-cap potentials}
\author[P. B. Jones]{P. B. Jones\thanks{E-mail:
peter.jones@physics.ox.ac.uk}\\
University of Oxford, Department of Physics, Denys Wilkinson building,
Keble Road, Oxford OX1 3RH, England\\}
\begin{document}

\date{ }

\pagerange{\pageref{}--\pageref{}} \pubyear{}

\maketitle

\label{firstpage}

\begin{abstract}

Development of the ion-proton pulsar model extends it to the limit of large unscreened polar-cap potentials, for example, as in the Vela pulsar, in which ion charges differ only by small increments from their complete screening values.  It is shown that the atomic number $Z$ of an ion following its passage from the canonical $Z_{0} = 26$ value through the electromagnetic shower region to the surface is not necessarily time-independent but can vary between fixed limits in an irregular or quasi-periodic way in a characteristic time of the order of $10^{4}$ s.  Thus at a certain $Z$ the system may transition to an unstable state of higher electric potential and it is argued that this is the physical basis for mode-changes, long-term nulls, periodic or otherwise. The model requires an orientation of magnetic dipole moment relative to rotational spin giving a positive corotational charge density.  Success of the model would fix the particle composition of the remaining parts of the magnetosphere, including the Y-point and is therefore relevant to X-ray and $\gamma$-ray emission processes.

\end{abstract}


\section{Introduction}

Radio pulsars may exhibit complex patterns of emission.  Mode-changes, nulls and sub-pulse drift have been observed for many years but recent work has added periodic phase-stationary amplitude fluctuations to their number (Yan et al 2019; Basu, Mitra \& Melikidze 2019). These observations are of normal pulsars though Mahajan et al (2018) had previously observed periodic mode-changes in the millisecond pulsar (MSP) B1957+20 whose period of 1.6 ms allowed their unambiguous definition.  Radio emission in general is greatly complex:  many pulsars exhibit only stochastic behaviour that might be expected in radiation whose source is the decay of plasma turbulence, others have one or more of the phenomena listed above.  Individual phenomena can present differently in different pulsars. The complexity is such that it would not be surprising if new phenomena remain to be identified.

These observations are of interest because they are the sole guide to the structure of the neutron-star open magnetosphere and should allow the inference of its plasma composition. 

In any work on pulsar emission it is essential to divide the various components into two classes: polar-cap and light-cylinder sources.  Incoherent emissions fall into the latter  as do some coherent emissions.  It has been argued previously (Jones 2016) that circular polarization observed at high resolution to be slowly varying as a function of longitude is an indication of a polar-cap source from which radiation passes to the observer through little birefringence.   (Except that transport of linear polarization through a non-uniform electron-positron plasma can lead to the production of a small circular component: see Beskin \& Philippov 2012; Petrova 2016.) Light-cylinder sources do not display this and are not within the scope of this paper.  Recent developments in this latter area, based on magnetic re-connection in the equatorial current sheet beyond the Y-point, have been described by Lyubarsky (2019) and Philippov et al (2019).

Study of polar-cap physics in relation to observed pulsar phenomena was greatly stimulated by the seminal paper of Ruderman \& Sutherland (1975) which assumed positive polar-cap corotational charge density and an ion work function large enough to bind ions at polar-cap temperatures.  But its publication preceded the discovery of MSP and of the $8.5$ s pulsar J2144-3933 which appear incompatible with its assumption of polar-cap pair creation.  However, we believe that its insight, that condensed-matter degrees of freedom must be involved, is correct and possible only for neutron stars with positive polar-cap charge density.

The ion-proton polar cap with positive corotational charge density is a model open magnetosphere with the merit of supporting a common source of coherent radio emission in normal pulsars, millisecond pulsars (MSP) and long-period pulsars such as J2144-3933.  It also shows how nulls occur with varying rotation period and whole-surface temperature (Jones 2020).  In these regards it has advantages over the canonical electron-positron pair models for which it is not even possible to show that pair creation of adequate multiplicity occurs in the majority of pulsars (Hibschman \& Arons 2001; Harding \& Muslimov 2002) unless special assumptions are made about multipole fields, which are implausible in the MSP and in the 8.5 s pulsar J2144-3933 (Young, Manchester \& Johnston 1999).  These latter authors questioned the electron-positron models as had Weatherall \& Eilek (1997).  

J2144-3933 is of particular interest because its profile full-width at half maximum and polarization had been observed (Manchester, Han \& Qiao 1998) before the rotation period was correctly assigned and found to be unremarkable as of the general population.  Similar findings have been published for the MSP (see Jenet et al 1998; Kramer et al 1999).  It is a fair conclusion that all classes of pulsar have a common polar-cap emission mechanism.

This paper attempts to relate this complexity as a whole to the ion-proton polar-cap model.  It was initially motivated by the B$\rightleftharpoons$Q transitions seen in B0823+26 (Hermsen et al 2018, Basu \& Mitra 2019).  Specifically, the profiles of the B and Q-modes seen by Basu \& Mitra at 320 MHz are remarkably similar to examples of variations in null structures that appear in solutions of the ion-proton model if either the unscreened acceleration potential or the whole-surface neutron-star temperature is varied.  Sudden temperature changes can be excluded.  Thus we are prompted to associate B$\rightleftharpoons$Q mode-changes with potential changes at irregular intervals of the order of $10^{2} - 10^{4}$ s above the part of the polar cap that contributes radiation to the observer line of sight.  The problem is then to find a physical mechanism for bistability or potential fluctuations on such time-scales, also to understand how the shorter time-scales of the phase-stationary periodic amplitudes appear.       This is described in Section 2.  Section 3 relates the results of Section 2 with observation.  Our conclusions are summarized in Section 4.

\section{Ion-proton model}

The ion-proton polar cap as studied so far in the approximation of azimuthal symmetry about the magnetic axis describes the formation of a two-beam system, ions and protons having velocity distributions well approximated on any given flux line by $\delta$-functions that are well separated owing to their different charge-to-mass ratios. Hence a longitudinal or quasi-longitudinal Langmuir mode exists with amplitude growth rate $\exp(\Lambda)$ sufficiently large for the formation of non-linearity and turbulence.  Whilst the longitudinal mode does not couple with the radiation field, quasi-longitudinal modes are capable of doing so (Asseo, Pelletier \& Sol 1990); more significantly, consideration of charge fluctuations in a plane perpendicular to the mode wave-vector suggests that coupling should be expected within turbulence generated by a longitudinal velocity field which is non-uniform over the polar cap. Non-uniformity is a necessary consequence of an electric field boundary condition on the surface separating the open from other sectors of the magnetosphere. At any instant, the plane would contain charge fluctuations without unity of phase as opposed to a strictly one-dimensional mode in which charge fluctuations in such a plane would be in phase.  Transverse electric field components would be present: the system would be equivalent to a disordered array of dipole moments perpendicular to the wave-vector. Details of the creation of a radio spectrum lie beyond the scope of this paper.  Subsection 2.1 summarizes the model of Jones (2020) and following Subsections examine some of its more important limitations.

The open sector charge density $\rho$ is not outstandingly controversial (Contopoulos 2016: it is the sector in which magnetic flux passes from the polar cap to the light cylinder without crossing the null charge density surface $\rho = 0$).  The corotational charge density $\rho_{GJ}$ is likely to be close to the CKF charge density (Goldreich \& Julian 1969; Contopoulos, Kazanas \& Fendt 1999) derived from the force-free condition with an equatorial current sheet.

 Here, the specific assumption is that at a small altitude $h \ll u_{0}$, where $u_{0}$ is the polar-cap radius, $\rho(h) = \rho_{GJ}(h)$ and that for altitudes $z > h$, a small value of $\rho_{GJ}(z) - \rho(z) > 0$ is maintained through ion photoelectric transitions.  But we must acknowledge that this is at variance with those authors who consider that the polar-cap current is determined not by local electric-field screening but by magnetospheric structure as a whole (see Mestel et al 1985; also Beloborodov 2008), with the consequence that the
current density $j(z) > j_{GJ}(z)$ is possible above the polar cap.  Timokhin \& Arons (2013) have argued that this leads to intermittency of $j$ on time-scales $\ll u_{0}/c$,  generating potential differences large enough to produce electron-positron pairs by curvature radiation and single-photon conversion.  Thus the conditions $j > j_{GJ}$ and $\rho = \rho_{GJ}$ can both be satisfied averaged over time intervals of the order of $u_{0}/c$. A space-charge-limited flow boundary condition at the neutron-star surface is assumed in this work but a positive corotational charge density is not considered.  In that case, a substantial fraction of the pair energy would be in the reverse electron flux resulting in a large proton production rate.  It is likely that a large parallel electric field above the polar cap would be screened by ions or protons within a single transit time, an order of magnitude more rapidly than by the growth of pair formation that might occur through the presence of a low level of leptons.
Recent numerical simulations (see, for example, Philippov, Spitkovsky \& Cerutti 2015) lend some support for the fixed current hypothesis, but owing to the assumption, for technical convenience, of a fictional large neutron-star radius, do not place adequate weight on regions where large screening adjustments would be necessary locally in the absence of pair creation. That the inductance of the magnetosphere open sector viewed as a single conductor is at most a logarithmic function of altitude does not affect the ion-proton model in which there is very little current variation.  We do not believe that these questions have been resolved at the present time.

One-dimensional models of the polar cap (Mestel et al 1985, Beloborodov 2008) also predict that the charge-density distribution for a pure proton or ion plasma at $j < j_{GJ}$ may exhibit fine structure characteristic of the Debye length.  But these models are valid only for $z \ll u_{0}$.  The parameters of the ion-proton model can be assumed to be time-averages over intervals short compared with the proton diffusion time $\tau_{p}$ but extremely long in relation to Debye lengths.
Consequently, we expect any fine structure in charge density to be removed by the presence of two particle types and non-uniformity of potential owing to failure of the one-dimensional approximation.  Any residual fine structure might even expedite nucleation of the Langmuir mode. 

For finite angle $\psi$ between the local magnetic flux ${\bf B}$  and the rotation spin ${\bf \Omega}$, the magnetosphere above the polar cap cannot be symmetric about the magnetic axis.  Its precise shape at the neutron-star surface depends on the form of the magnetic flux density at the light cylinder.  But a limited idea can be obtained by consideration of the variations in $\rho - \rho_{GJ}$, the source of the acceleration field ${\bf E}_{\parallel}$.  The unscreened potential $\Phi_{ref}$ is not the vacuum potential but is defined here as that given by a charge density on any flux line that is determined by the condition $\rho = \rho_{GJ}(h)$ at a fixed low altitude $h$ above the polar cap surface. Hence regions in which $\rho - \rho_{GJ} < 0$ contribute to a finite acceleration field.  On lines with ${\bf E}_{\parallel}$ such that acceleration to the light cylinder is not possible there is stasis with $\rho = \rho_{GJ}$.   We assume that the unscreened potential is $\Phi_{ref} = 0$ on any flux line in the surface separating the open from other sectors of the magnetosphere. Under the assumption that $\rho$ consists of a single particle sign, is subject to a continuity condition, and that $\rho_{GJ} \propto -\cos\psi$, it is apparent that $\Phi_{ref}$ on any flux line in the open sector is principally a function of $\pi - \psi - \beta$ measured for that line on the polar-cap surface.  Here, $\beta$ is the observer line-of-sight angle relative to ${\bf \Omega}$.  Lines that bend towards ${\bf \Omega}$ with increasing radius $\eta$ (in units of the neutron-star radius) have the greatest acceleration potential: those bending in the opposite sense towards the null surface have lower $\Phi_{ref}$.

An asymmetric $\Phi_{ref}$ indicates that only a band of the polar cap with $\pi - \psi - \beta$ values within limits can support pulsar emission as described in Subsection 2.1.  In the limit of maximum $\beta$, $\Phi_{ref}$ is small and essentially indeterminate as lines become close to the sector in which lines approach the Y-point (see Contopoulos 2016). An important consequence of the asymmetry is that a polar cap could, in principle, simultaneously support two different plasma modes.

Pair creation by curvature radiation is not possible, without special assumptions, in a large fraction of pulsars.  Photo-electric transitions in ions accelerated through the whole-surface blackbody radiation field produce reverse-accelerated electrons and  electromagnetic showers in the compact neutron-star atmosphere which, apart from proton creation, could be a significant source of pairs above the polar cap in circumstances where $B_{12}$ and $\Phi_{ref}$ are large enough to support inverse Compton scattering.   The major source is likely to be (n,$\gamma$) reactions, the $\gamma$-rays moving outward through the neutron-star surface and producing pairs by single-photon conversion: the neutrons are produced in showers at approximately the same rate as protons.  As a conceivable source of self-sustaining electron-positron pair creation, this was studied by Jones (1977, 1979): approximately $1 - 2$ pairs are produced per $10^{3}$ GeV electron shower energy.  Thus it would be limited to pulsars with $B_{12} > 2$ and large values of $\Phi_{ref}$ and could not be applicable to MSP or to J2144-3933 and many other normal pulsars.  Also, it is now thought that there are current sheets from the polar-cap periphery to the vicinity of the Y-point.  An inward particle flux is likely to be the origin of the high temperatures ($> 10^{6}$ K) observed in small areas of the neutron-star surface: it is also likely to be a further source of (n,$\gamma$) reactions.

Thus in the limit of large $\Phi_{ref}$ an inverse Compton scattering process may support self-sustaining pair creation.  A previous publication (Jones 2018) has described this state for a symmetric polar cap.  Outward plasma consists of protons, and small fluxes of positrons and low-energy secondary pairs.  There are no ions with $Z \geq Z_{min}$, where $Z_{min}$ is the least atomic number of ions in local thermodynamic equilibrium at polar-cap temperatures that are not completely stripped of electrons. Hence there is no contribution from ions to the reverse-electron energy flux.  The development of an atmosphere of protons guarantees the stability of this phase, which could be long-lived.

\subsection{The symmetric polar cap}

The model applies to positive polar-cap charge densities in which ions of mean atomic number $Z \geq Z_{min}$, charge $Z_{h}$, are accelerated through the whole-surface blackbody radiation field and thereby photo-ionized to a final charge $Z_{\infty}$.  The reverse flux of electrons creates electromagnetic showers and a proton flux that, with a diffusion time-delay $\tau_{p}$ of the order of $1$ s, moves under the influence of a small electric field to the top of the neutron-star atmosphere.  Evaluation of the model uses a cellular polar cap in which Poisson's equation relates the partially screened potential $\Phi$ to $Z_{\infty}$ over the whole polar cap, but in each cell the local values of $Z_{\infty}$, $\Phi$ and of the reverse electron energy $\epsilon$ per ion are in known relationships with each other.  Dependence on photoionization means that the relations are also functions of whole-surface neutron-star temperatures.  Owing to their higher charge-to-mass ratio, protons are accelerated in preference to ions.  Hence, in general, the outward flux of accelerated particles is of protons, with ions making up any deficit in charge density to the Goldreich-Julian value.  We refer to Jones (2010, 2020) for more complete details.

Development of the outgoing plasma consists firstly of the growth of the Langmuir mode within a radius of the order of $\eta = 5 - 10$ ($\eta$ is the polar radius in units of the neutron-star radius $R$).  Adequate values of the amplitude growth-rate exponent $\Lambda$ require moderate Lorentz factors, typically $\gamma_{p} < 50$ for protons.  But this is quite possible because the acceleration field is almost completely screened by the reverse electron flux.  In model solutions, the Poisson equation is satisfied at each time step, the polar cap having been divided into 91 discrete cells of equal area.  The ion and proton charge densities are represented by coefficients $\alpha_{Z} + \alpha_{p} = 1$ at the low altitude $h$ where motion has become relativistic.  These are calculated for each cell and for each time-step.

Mode growth requires two beams.  Thus a cell with $\alpha_{p} = 1$ at any step does not contribute to turbulence and radio emission during that step.  The growth exponent $\Lambda \propto \gamma_{p}^{-3/2}$ approximately, so that a modest change in $\gamma_{p}$ can produce a turbulence cut-off.  Thus necessary and sufficient conditions for radio emission are finite $\alpha_{Z}$ and $\alpha_{p}$  and Lorentz factors below a critical value, $\gamma_{p} < \gamma_{c}$ within an area large enough to provide transverse dimensions adequate for mode growth.  Failure of these conditions on those areas of the polar cap that contribute to the observer line of sight produces a null.  The model demonstrates the increasing frequency and length of nulls as the neutron-star whole-surface temperature decreases, as it does with age.  It also shows that an increase in the unscreened acceleration potential has the same effect.  The output of the model is chaotic from step to step but in a quasi-stable manner as is observed in pulsars.  But nulls or mode-changes as long as $10^{4}$ periods in length are not a feature of existing solutions of the model.

\subsection{Large unscreened potentials}

Solution of the symmetric ion-proton model is difficult in the limit of large unscreened potentials, for example, as in the Vela pulsar.  The problem is that the partially screened potential $\Phi$ is several orders of magnitude smaller than $\Phi_{ref}$.
Therefore the final ion charge $Z_{\infty}$ must be extremely close to the complete screening value and so determines the local $\Phi$ and the energy of the reverse electron flux, $\epsilon$ per ion. Because $Z_{\infty}$ is so close to the complete screening value (for which $E_{\parallel} = 0$), Poisson's equation is not a useful relation here. A further problem is that the limit $Z_{\infty} \leq Z$ may be encountered in one or more of the polar-cap cells.(For reference purposes, the maximum potential energy for a symmetrical polar cap is $\approx 1.2\times 10^{3} B_{12}P^{-2}$ GeV.)

In the general formulation of the model, the plasma charge density evolves by photo-ionization from its value at the low fixed altitude $h$,
\begin{eqnarray}
\rho(h) = \rho_{GJ}(h) = {\rm e}N_{Z}(2Z_{h} - Z_{\infty} + K)\nonumber \\
 = (\alpha_{Z} + \alpha_{p})\rho_{GJ}(h),
\end{eqnarray}
in which $K$ is the number of protons per ion and $N_{Z}$ is the ion number density,
to its value at infinity, assumed here in model evaluation to be a radius $\eta = 4$,
\begin{eqnarray}
\rho_{\infty} = {\rm e}N_{Z}(Z_{\infty} + K)
\end{eqnarray}
in the process following closely the change in $\rho_{GJ}$ with altitude derived from the Lense-Thirring rotation (Beskin 1990, Muslimov \& Tsygan 1992).  The proton flux is determined by $\epsilon$,
\begin{eqnarray}
\alpha_{p}(t) = W_{p}\int^{t}_{-\infty}dt^{\prime}\frac{\epsilon(t^{\prime})
(1 - \alpha_{p}(t^{\prime}))f_{p}(t - t^{\prime})}{2Z_{h} - Z_{\infty}(t^{\prime})},
\end{eqnarray}
valid locally over the polar cap and in which the function $f_{p}$, normalized to unity, describes the diffusion time delay for protons from formation to arrival at the top of the neutron-star atmosphere.  The constant is $W_{p} = 0.2$ GeV$^{-1}$  (see equations (4) - (7) and  (9) - (11) of Jones 2020).

An approximation valid for large $\Phi_{ref}$ is that of placing $Z_{\infty}$ equal to its complete screening value for which $\Phi = 0$.  The actual value of $\Phi$ and of $\epsilon$ are obtained directly from the complete screening value of $Z_{\infty}$ using the temperature-dependent relations between $Z_{\infty}$, $\Phi$ and $\epsilon$. The relation between $\alpha_{p}$ and $Z_{\infty}$ at any instant is,
\begin{eqnarray}
\alpha_{p} = 1 - \frac{\kappa}{1 - \kappa}\frac{2Z_{h} - Z_{\infty}}
{2(Z_{\infty} - Z_{h})},
\end{eqnarray}
in which $\kappa = 0.15$ is the Lense-Thirring factor and $Z_{h}/(1 - \kappa) < Z_{\infty} < Z$. 
We assume this to be valid over the whole polar cap except for areas in which $\Phi_{ref}$ is small. Solutions of equations (3) and (4) exist for suitable $\epsilon$, assuming time-independence of all quantities, and are given by
\begin{eqnarray}
\epsilon W_{p} = \left(\frac{2 - \kappa}{\kappa}\right)Z_{\infty} - \frac{2}{\kappa}Z_{h},
\end{eqnarray}
bearing in mind that $\epsilon$ is a known function of $Z_{\infty}$ at a given whole-surface temperature (see the first paragraph of Section 2.1).
But for many intervals of $Z_{h}$ and $Z_{\infty}$ solutions do not exist; also it remains to find the conditions necessary for the stability of any solution by examining solutions $Z_{\infty} \rightarrow Z_{\infty} + \delta Z_{\infty}(t)$ in its vicinity.
Equation (3) can be re-expressed as
\begin{eqnarray}
\delta Z_{\infty}(t) = CW_{p}\int^{t}_{-\infty}dt^{\prime}\delta Z_{\infty}(t^{\prime})f_{p}(t - t^{\prime})
\end{eqnarray}
by the elimination of $\alpha_{p}$, where,
\begin{eqnarray}
C = \frac{1}{Z_{h}}\left((Z_{\infty} - Z_{h})\frac{\partial \epsilon}{\partial Z_{\infty}} - \epsilon \right)
\end{eqnarray}
expressed in time-independent values.  Examination of $\epsilon$ as a function  of $Z_{\infty} - Z_{h}$ shows that it increases rapidly. Assuming $\epsilon \propto (Z_{\infty} - Z_{h})^{n}$, it follows that $C = (n - 1)\epsilon/Z_{h}$.  This indicates that stability is possible only for small-$\epsilon$ solutions.Then for a fluctuation $\delta Z_{\infty}(t^{\prime}) = \delta Z_{\infty}(t_{0})g(t^{\prime} - t_{0})$ in which $g$ is normalized to unity, the condition for instability is $CW_{p} > 1$.

Equations (4) and (5) are no longer valid in the case of instability. The $\alpha_{p}$ given by equation (4) is subject to a maximum  $\alpha_{p}^{c} < 1$ for $Z_{\infty} = Z$. Instability of the complete screening approximation has the consequence that the acceleration potential promptly increases.  In the complete screening limit, Poissons's equation is not involved so that the stability condition applies locally and independently to any area of the polar cap. But instability in any area of the polar cap has the capacity, though dependent on their existing margins of stability, to cause instability in adjacent areas because, owing to the long-range electric potential, there is a local change in $\Phi$, in $Z_{\infty}$, and $\epsilon$.

The Vela pulsar has a maximum $\Phi_{ref} \approx 300$ TeV and the model of this section must be applicable.  With the parameter $W_{p} = 0.2$ GeV$^{-1}$, whole-surface temperature
$T_{s} = 5.8 \times 10^{5}$ K (\"{O}zel 2013),  $Z = 10$ and $Z_{h} = 8$, a time-independent solution satisfying equation (5) exists.  But variation of the parameters $Z$ and $Z_{h}$ indicates that its existence is parameter sensitive to the extent that given the uncertainties in these parameters and in electron binding energies and photoelectric cross sections, it is not possible to assert with confidence that the state exists.  In the event that no solution of equations (4) and (5) exists or any solution is not stable, the system is chaotic with $Z_{\infty}$ oscillating between the limits $Z_{h}/(1 - \kappa) < Z_{\infty} < Z$, the time scale being of the order of
$\tau_{p}$.  The potential $\Phi$ is also oscillatory and in young pulsars such as Vela is likely to become large enough for pair creation to exist in these short intervals, most probably through the inverse Compton scattering mentioned previously.  Time intervals in which $\Phi$ is small enough to permit functioning of the ion-proton mode could be non-existent.

\subsection{Instability of surface nuclear charge}

In this Subsection we consider the stable state defined in Subsection 2.2.  Here
$\epsilon$ and therefore $C$ are functions of $Z_{\infty}$ and $Z_{h}$ and hence of the surface atomic number $Z$.  In electromagnetic showers, the initial atomic number $Z_{0}$ of a nucleus is gradually reduced to $Z$ by giant dipole-state formation and decay by proton and neutron emission.  Almost all of this occurs at a depth of about $10$ radiation lengths where the track length of $15 -25$ MeV photons peaks sharply.
If there were no subsequent neutron capture and $\beta$-interactions, 
$Z_{0} = \langle Z \rangle + \langle K \rangle$ in which the averages are taken over a long time interval.  With complete neutron capture and $\beta$-equilibrium achieved before the original nucleus reaches the surface we should expect similarly, in terms of nuclear mass numbers, $A_{0} = \langle A \rangle + \langle K \rangle$, with $\langle A \rangle \approx 2\langle Z \rangle$.  The evolution from $Z_{0}$ to $Z$ is monotonic in the first case but not necessarily so in the second.  The true state lies between the two extremes but is difficult to estimate with any confidence.
Whatever the relation, it must be satisfied over a long period of time.

Instead, we shall assume $A_{0} = 2Z_{0}$, also $\langle A\rangle = 2\langle Z \rangle$ and adopt an intermediate state so that
\begin{eqnarray}
\langle K \rangle = (3/2)(Z_{0} - \langle Z \rangle)
\end{eqnarray}
is the number of protons released per nucleus $(A_{0},Z_{0})$.  We have
$\langle K \rangle = W_{p}\langle \epsilon \rangle$ for any interval of time large compared with $\tau_{p}$. 
Averaged over long time intervals equivalent to the loss from the surface of  more than a shower depth, which is of the order of $10 l_{r}$, the value of $\langle K \rangle$ must in the complete screening limit satisfy both equations (5) and (8). (Here $l_{r}$ is the radiation length defined for electromagnetic bremsstrahlung.)

The question is whether time-independent values of $\langle K \rangle$ and $Z$ exist which satisfy this condition.  Any arguments made here are to some extent vitiated because the value of $Z_{0}$ is
unknown and may even depend on the neutron-star formation process and on fall-back. The condition is,
\begin{eqnarray}
Z = Z_{0} - \frac{4}{3\kappa}(Z_{\infty} - Z_{h} + \frac{2}{3}Z_{\infty}.
\end{eqnarray}
 We assume the canonical value $Z_{0} = 26$.  Bearing in mind that for any value of $Z$, $Z_{h}$ and $Z_{\infty}$ are quite closely defined by considerations of ionization and screening, we see that time-independent solutions can exist only for small values of $Z_{\infty} - Z_{h}$.  The two physical processes concerned here are quite different but are connected on two different time-scales:  the time $\tau_{p}$ for proton diffusion to the top of the atmosphere, and the time interval $\tau_{Z}$ in which nuclei of charge $Z_{0}$ enter the electromagnetic shower region, undergo successive photo-disintegrations to charge $Z$ and move to the surface of the atmosphere.  The value of $Z$ at any instant determines the reverse-electron energy flux $\epsilon$ and therefore the value of $Z$ at a time $\tau_{Z}$ later.  If a time-independent state were to exist it must have $Z \geq Z_{min}$, where $Z_{min}$ is the smallest nuclear charge for which the local thermal equilibrium ion charge $Z_{h}$ has electrons that can be stripped by photo-ionization to provide screening.  Without screening, ions of $Z < Z_{min}$ leave freely and are accelerated to high Lorentz factors such that the Langmuir mode cannot grow to create turbulence (except in the $8.5$ s pulsar J2144-3933; Jones 2020). In these circumstances, there is no reason why a time-independent state should invariably exist but the long-term condition defined by equations (5) and (8) must be satisfied.  The most obvious way in which this could happen is an irregular or quasi-irregular alternation between different values of $Z$ which is possible in the region of the instability described in Subsection 2.2.  This would produce a two-phase system.  The natural time-scale for the existence of either phase is at least of the order of $t_{rl}$, the time in which at Goldreich-Julian density, one radiation length of matter leaves the neutron-star surface,
\begin{eqnarray}
t_{rl} = 2.1\times 10^{5}\left(\frac{P}{ZB_{12}\ln(12Z^{1/2}B_{12}^{-1/2}}\right)
\hspace{5mm} {\rm s}  
\end{eqnarray}
 in which the expression for the radiation length $l_{r}$ has been adjusted to a magnetic field of the order of $B_{12} = 1$ (see Jones 2010).  This is of the order of $10^{4}$ s for values of $B_{12}$ and $P$ considered here and would be an estimate of the time for  movement of an ion through the equivalent of one radiation length assuming that there is no relative diffusion of different ion species.  Therefore $\tau_{Z} \approx 10 t_{rl}$ as would be the case in a solid phase, ordered or otherwise.  But an atmosphere in local thermodynamic equilibrium exists, though of unknown depth, above a density discontinuity  and a solid or possibly liquid phase.  Relative diffusion of ion species is in part entropy-driven but there is a potential gradient that fractionates according to ion charge-to-mass ratio.  High-$Z$ ions in local thermodynamic equilibrium have the lowest charge- to-mass ratio owing to the magnitudes of electron binding energies compared with the thermal quantum.  Low mass-number ions of $A = 2Z$ diffuse preferentially to the top of the atmosphere immediately below any proton component that is present.  Thus the time between an ion of atomic number $Z_{0}$ entering the shower maximum region and its residual ion $Z$ reaching the neutron-star atmosphere could be shorter than $10t_{rl}$ but is certainly long compared with $\tau_{p}$.
 
\section{Potential bistability or fluctuations}

Application of the ion-proton model leads at once to the view that radio-frequency emission is a two-stage process.  The longitudinal 
Langmuir mode does not couple directly with the radiation field  so that growth to non-linearity and turbulence is needed.  The growth rate $\exp(\Lambda)$, with $\Lambda \propto \gamma^{-3/2}_{p}$ is numerically large.  Consequently a modest fractional change in $\gamma_{p}$, the result of a similar change in polar-cap potential $\Phi$, can produce a sharp cut-off in turbulence formation and hence observable radiation.

A working hypothesis is that all complex observed phenomena reflect no more than modest changes in $\Phi$.  An example is provided by the B$\rightleftharpoons$Q
mode-changes in B0823+26 (Hermsen et al 2018; Basu \& Mitra 2019).  The radio-bright B-mode has unremarkable radio emission with a small fraction of short nulls.  The Q-mode radio profiles appear to depend on the observing occasion but consist of short burst lengths and long nulls of typically $10^{2}$ periods.  As noted in Section 1, the B$\rightleftharpoons$Q  transitions are remarkably similar to the change in ion-proton model emissions that occurs if the unscreened potential $\Phi_{ref}$ is increased by a factor of about $1.2$ (see Jones 2020: Table 2) leading to a rather small increase in $\Phi$.  In the model calculations, a larger change suppresses radio emission completely.

The second factor that appears significant is the dependence of unscreened potential $\Phi_{ref}$ on position within the polar cap, specifically on $\pi - \psi - \beta$, where $\psi$ is here the angle subtended at the rotation axis by a flux line, for which arguments were given in Section 2.  The existence of a wide interval of $\Phi_{ref}$ prompts the following question.  Can an open magnetosphere consisting of two distinct states characterized by quite different intervals of values of $\Phi$ exist above a polar cap?  For example, can the inverse Compton scattering mode described in Section 2 be supportable in a region of high $\Phi_{ref}$ and stable to the extent that its continued existence depends only on parameters within its own sector but is independent of physical conditions in the small $\Phi_{ref}$ sector?  We would argue in the affirmative, but note that a potential change in a specific sector must always affect the potential in any adjoining sector owing to the long range of the Coulomb force.

Radiation parallel with the observer line-of-sight is emitted from a finite-width band in the values of $\pi - \psi - \beta$ owing to the moderate Lorentz factors of protons and ions leading to significant finite emission angles.  Therefore, the potential $\Phi$ within this band can be changed discontinuously to the modest extent needed to produce a null or mode-change by a potential change outside this band in a polar-cap region not visible to an observer and possibly not emitting radio-frequency power.

The existence of this band is also relevant to all pulsars with large unscreened potentials, such as Vela and many MSP, which also exhibit uncomplicated emission consistent with an ion-proton polar-cap.  Section 2.2 indicates that stable-state solutions of equation (5) are quite unlikely to exist over the whole polar cap.  But these pulsars have large polar caps and the possibility of wide bands of small $\Phi_{ref}$ and $\gamma_{p}$ so that their observability is not compromised.  The surface atomic number time constant $\tau_{Z}$ described in Section 2.3, with the proton diffusion time constant $\tau_{p}$, form the basis for understanding complex emission profiles in the ion-proton model.

\subsection{Short time-scales}

Some evaluations of the model have been published previously (Jones 2020) and have shown how the existence of $\tau_{p}$ gives an understanding of stochastic behaviour with short null lengths $ \leq 10^{2}$ s in normal pulsars.  Recently, Mahajan et al (2018) have observed short-lived mode-changes with an average length of $1.7$ s in MSP B1957+20.  Here the main pulse is phase-aligned with the $\gamma$-emission, has a very small mode-change and a low level of circular polarization.  The inter-pulse has substantial mode-changes and circular polarization.  On this basis we assign the inter-pulse to polar-cap emission and the main to light-cylinder emission. The order of magnitude of the maximum $\Phi_{ref}$ for this pulsar is $80$ TeV indicating that a quasi-periodic potential fluctuation on the time-scale of $\tau_{p}$ in a high-$\Phi_{ref}$ region could be the explanation for the observed $1.7$ s interval. As these authors note, it would be interesting to examine more MSP for this phenomenon.

More recently, a number of authors have observed a new phenomenon, phase-stationary periodic mode-changing (Yan et al 2019; Basu, Mitra \& Melikidze 2019).  Basu et al published a catalogue of $18$  instances in normal pulsars and have presented an interesting comparison with periodic nulls.  Periods have been found from longitude-resolved frequency spectra (LRFS) and, expressed in seconds, $15$ of the $18$ pulsars in the catalogue lie in a moderately compact interval of $2.7 - 16.3$ s.  Yan et al found a period of $2.1$ s in the young normal pulsar J1048-5832 also derived from LRFS.

These authors compare periodic mode-changing with periodic nulls.  Basu et al also catalogue $29$ normal pulsars exhibiting periodic nulls.  But there are differences between the average periods of the two sets, also between their average neutron-star kinetic life-times.  Here, we express periods in seconds and note that the periodic nulls lie in the interval $7-266$ s, an order of magnitude larger than the mode-changing periods.  The mean lives (Manchester et al 2005) are $3.3\times 10^{7}$ yr for periodic nulls, but only $6.2\times 10^{6}$ yr for the periodic mode-changing pulsars.

Except in a few cases, the ionization in the ion-proton model is caused by the whole-surface blackbody field and we can be certain that its temperature $T_{s}$ decreases with age.  Periodic nulling pulsars have a compact distribution of $\Phi_{ref}$, all except two being in the interval $0.5 - 2.2$ TeV.  The quoted periods are consistent with those predicted at these values of $\Phi_{ref}$ for $T_{s}$ in a compact band below $3\times 10^{5}$ K (Jones 2020).  The younger and warmer periodic mode-changers have shorter periodicities as expected.

The explanation is that both nulls and mode-changes are caused by fluctuations in potential above the observable band of the polar cap.  In the case of nulls, the primary change may be above the band, whereas in mode-changes it is likely to be outside the observable band in a region of larger $\Phi_{ref}$ with only secondary changes above the band.  Periodicity is the only difference from the mode-change and null mechanisms described previously (Jones 2018, 2020).  Pulse sequences observed by Yan et al have a well-defined LRFS frequency but do not appear as a high-Q system.  The existence of the proton diffusion time-constant $\tau_{p}$ is the basis for the periodicity.  In any element of polar-cap area it governs the sequences in which an interval of ion emission with large $\epsilon$ is followed by one in which protons diffuse to the top of the neutron-star atmosphere at a rate so high that ion emission and therefore local electric field screening is removed completely above that area with a consequent upward fluctuation in potential $\Phi$.  

Periodicity defined by $\tau_{p}$ is therefore very plausible but unfortunately the cellular evaluation of the model described by Jones (2020) is not suitable for its demonstration. This is also true in the case of subpulse drift, a phenomenon which has been described previously (Jones 2014) but only in outline within the ion-proton model.  With the exception of B1957+20, the pulsars considered here are all normal.  It would be interesting to see if there is further evidence in MSP for the presence of time-constants of the order of $\tau_{p}$.

\subsection{Longer time-scales}

Nulls and mode-changes with time-intervals of the order of $10^{2-4}$ s have been observed in pulsars which are also X-ray emitters.  Apart from B0823+26, this is true for B0943+10 (Hermsen et al 2013, Bilous et al 2014, Mereghetti et al 2016, Rigoselli et al 2019).  The B-mode of B0823+26 emits X-rays principally below $10$ keV which, in the Q-mode are much reduced in intensity.  B0943+10 emits similarly at radio frequencies but with an X-ray intensity increased in the Q-mode.  In B1822-09 (Hermsen et al) X-ray emission is independent of B$\rightleftharpoons$Q mode-changes that are seen at radio frequencies.  Coincidence between radio and X-ray mode-changes indicates a link between magnetosphere changes above the polar cap and in the vicinity of the Y-point, but one that is not universal in detail.  Its origin could be the current flow believed to extend from the boundary of the polar cap to the Y-point (Contopoulos 2016, Contopoulos \& Stefanou 2019).  But this lies far beyond the scope of the present paper.

We believe the bistability in the surface atomic number $Z$ described in Section 2.3 to be the source of the longer null and mode-change time-scales. Secular change in $Z$ on time-scales $\tau_{Z}$ is a mechanism whereby the instability condition described in Section 2.2 can be activated.  In younger pulsars, polar-cap regions of high $\Phi$ approaching $\Phi_{ref}$ are possible sources of self-sustaining pair creation through inverse Compton scattering.  If there were observable consequences of this, they would be seen during radio-quiet intervals.

A small number of pulsars are known to exhibit null lengths of $10^{7-8}$ s, some orders of magnitude larger than $\tau_{Z}$.  It is tempting to ascribe these to prolonged self-sustaining inverse Compton scattering modes.  Nulls long enough to enable spin-down torque measurement are seen in B1931+24 (Kramer et al 2006), J1832+0029 (Lorimer et al 2012), J1841-0500 (Camilo et al 2012), J1910+0157 and J1929+1357 (Lyne et al 2017).Values of $\Phi_{ref}$ for these pulsars are of the same order as for B0823+26 and B0943+10.  The spin-down torque is approximately halved in the null state which, in the ion-proton model, could have limited pair creation by inverse Compton scattering for the values of $\Phi_{ref}$ concerned. Does loading of the electromagnetic field at and beyond the light cylinder by a multiplicity of electron-positron pairs increase or decrease the spin-down torque? The answer is not obvious to the author.

\section{Conclusions}

This and a previous paper (Jones 2020) are attempts to show that an ion-proton open magnetosphere exists in almost all observable radio pulsars.  The primary assumption made is that the polar-cap corotational charge density is positive.  The characteristics of the model then follow directly from well-established cross-sections and processes in nuclear and radiation physics.  The model is not particularly simple in application and depends on neutron-star parameters that are not well-known, particularly the whole-surface temperature $T_{s}$ and surface atomic number $Z_{0}$.  Evidence on $T_{s}$ is now emerging through ultra-violet measurements on the Rayleigh part of the Planck spectrum, but our use of the canonical value $Z_{0}  = 26$ is no more than a convenient assumption, neglecting details of the formation process and possible fall-back.  Whilst the model may not appeal owing to its complexity, this is to be  expected if condensed-matter degrees of freedom are involved non-trivially.

Our case is that the model does provide a physical basis for understanding the complexities of radio emission and hence the physics of the polar cap and open magnetosphere, which otherwise is an unsolved problem.  Establishing the polar-cap  corotational charge density is itself important because it determines the particle composition of the remaining sectors of the magnetosphere including the Y-point.

It might be complained that we have not dealt satisfactorily with the existence of periodicity.  But we have stressed that the appearance of the phenomena considered here is by no means universal nor is the manner in which they present in any given pulsar.  It is likely that specific intervals of parameter values are necessary for observable periodicity: non-uniformity of polar-cap potential or of $Z$ and $Z_{h}$  also complicate the problem.  Whilst a broad physical understanding of these problems is possible and has been attempted here, investigations specific to any given pulsar will prove much more difficult. 

\section*{Acknowlegments}

It is a pleasure to thank the anonymous referee for some very helpful comments, particularly concerning the precise nature of the polar-cap current-field relation.

\bsp

\label{lastpage}

\end{document}